\documentclass[sigconf,10pt]{acmart}
\usepackage[]{graphicx}
\usepackage{tikz}
\usepackage{url,balance}
\usepackage{enumitem,color}
\usepackage{colortbl}
\definecolor{Gray}{gray}{0.9}
\def\etal{\emph{et al. }}
\pdfoutput=1

\title{How to Design Browser Security and Privacy Alerts}
\author{Lynsay A. Shepherd}

\affiliation{%
School of Design and Informatics \\Abertay University 
\\Dundee, United Kingdom
}
\email{lynsay.shepherd@abertay.ac.uk}

\author{Karen Renaud}

\affiliation{%
School of Design and Informatics \\Abertay University 
\\Dundee, United Kingdom
}
\email{k.renaud@abertay.ac.uk}

\date{}

\acmConference[AISB'18]{}{5th April 2018}{AISB 2018 Symposium}
\acmYear{2018}

\begin{document}

\begin{abstract}
It is important to design browser security and privacy alerts so as to maximise their value to the end user, and their  efficacy in terms of communicating risk. We derived a list of design guidelines from the research literature by carrying out a systematic review. We analysed the papers both quantitatively and quantitatively to arrive at a comprehensive set of guidelines.  Our findings aim to to provide designers and developers with guidance as to how to construct privacy and security alerts.  We conclude by providing  an alert template, and highlighting its adherence to the derived guidelines.
\end{abstract}

\maketitle

\section{Introduction}
It is non-trivial to design effective alerts  in the security and privacy domain. 

Browser designers do their best to inform users about security-related aspects as they surf the web. Owing to the number of potential pitfalls, this means 
end users can effectively be bombarded with security alerts  \cite{acar2016you}, and users often ignore them \cite{kim2009habituation,anderson2014users}. Developers sometimes make unfounded assumptions about the background knowledge of alert recipients \cite{kauer2012not} making them incomprehensible. 

Privacy alerts are not perfect either \cite{larose2007promoting}. 
Users are often overwhelmed by these alerts because there are too many, \cite{egelman2008you} or because they do not know what actions to take as a consequence \cite{shankar2006doppelganger}.

This diminishes the impact of alerts, and leaves users vulnerable to unknowingly carrying out actions which will compromise their privacy or security.

Traditional usability guidelines cannot necessarily be used ``as-is'' in the security and privacy context. This is because neither privacy nor security are the end user's primary task \cite{balebako2013little,kelley2009designing}. Alerts interrupt the user's pursuit of their primary goal and are thus often perceived to be a nuisance  \cite{albrechtsen2007qualitative}.
We therefore need  specific guidelines to inform alert design in the security and privacy context. 

Much has been written about alert design, as can be seen from the following section. Yet one can hardly expect busy deadline-driven software engineers, the very people who are producing these alerts, to keep up with the latest research in the area. 
 
We therefore carried out a systematic literature review in order to consolidate all the published guidelines into one coherent list (Section \ref{analysis}). Previously, Bauer \etal \cite{Bauer13} presented a list of warning design guidelines in 2013.  Our work provides an updated, more comprehensive, list of guidelines, which are specifically tailored towards  browser-based alerts. 

Having derived a comprehensive set of deadlines, we realised that merely providing a  list of guidelines is not an optimal way of supporting designers.
Luger and Rodden \cite{luger2014value}  argue that such lists of guidelines are unlikely to be followed in the pressured environment of software development and design. Moreover, some of the published guidelines conflict \cite{DBLP:conf/hcse/MasipMWPGO12}, which is unhelpful.  

To make our consolidated guidelines as helpful as possible, we decided to convey the {\em spirit}, rather than the letter, of the guidelines in the form an example alert template (Section \ref{examples}). This conveys  the ``how'' of alert design, rather than the ``what'', as encapsulated in a linear set of alert guidelines.

Future work is explored in Section \ref{futurework}, and we conclude  in Section \ref{conc}.

\section{Informing End Users}
First we clarify the nomenclature used in this paper. We then provide an overview of the human in the loop model of human information processing. We conclude by explaining the difference between the foundational security and privacy concepts. 

\subsection{Nomenclature}
We investigated guidelines that inform the design of  warnings, alerts, notifications, prompts or provision of feedback. The underlying concept is the same: provision of important information to an end user that the system considers he or she should be apprised of. We shall use the term `alert' as a unifying term to represent  all the terminologies used by papers cited in this paper.

\subsection{Human Information Processing}
Wogalter and Mayhorn \cite{Wogalter2017} explain that warnings (what we call alerts) are a type of risk communication. Wogalter \cite{wogalter1999factors} explains that warnings have two purposes: (1) communicate information, and (2) reduce unwise behaviours. To achieve these aims the warnings have to be designed carefully.  
The Web Content Accessibility Guidelines{\footnote{\url{https://www.w3.org/TR/WCAG21/}}} can also be applied to alerts \cite{almeidamerging} i.e. that they should be  perceivable, operable, understandable and  robust. 

Shannon \cite{shannon2001mathematical} and Lasswell \cite{lasswell1948structure} both proposed models of human communication which help us to understand how humans process alerts. 

Wogalter, DeJoy, and Laughery \cite{wogalter1999organizing} developed the C-HIP model in the context of warning research. Their model builds on the work of Shannon and Lasswell and can be considered to be somewhat unrealistic because it does not include a noise component. In a world of noisy communication such a model is incomplete. Cranor \cite{cranor2008framework} proposed a human-in-the-loop framework which is more comprehensive and reflects the factors impacting communications in the  context of privacy and security alerts.

\begin{figure}[ht]
	\centering
        \includegraphics[width=\columnwidth]{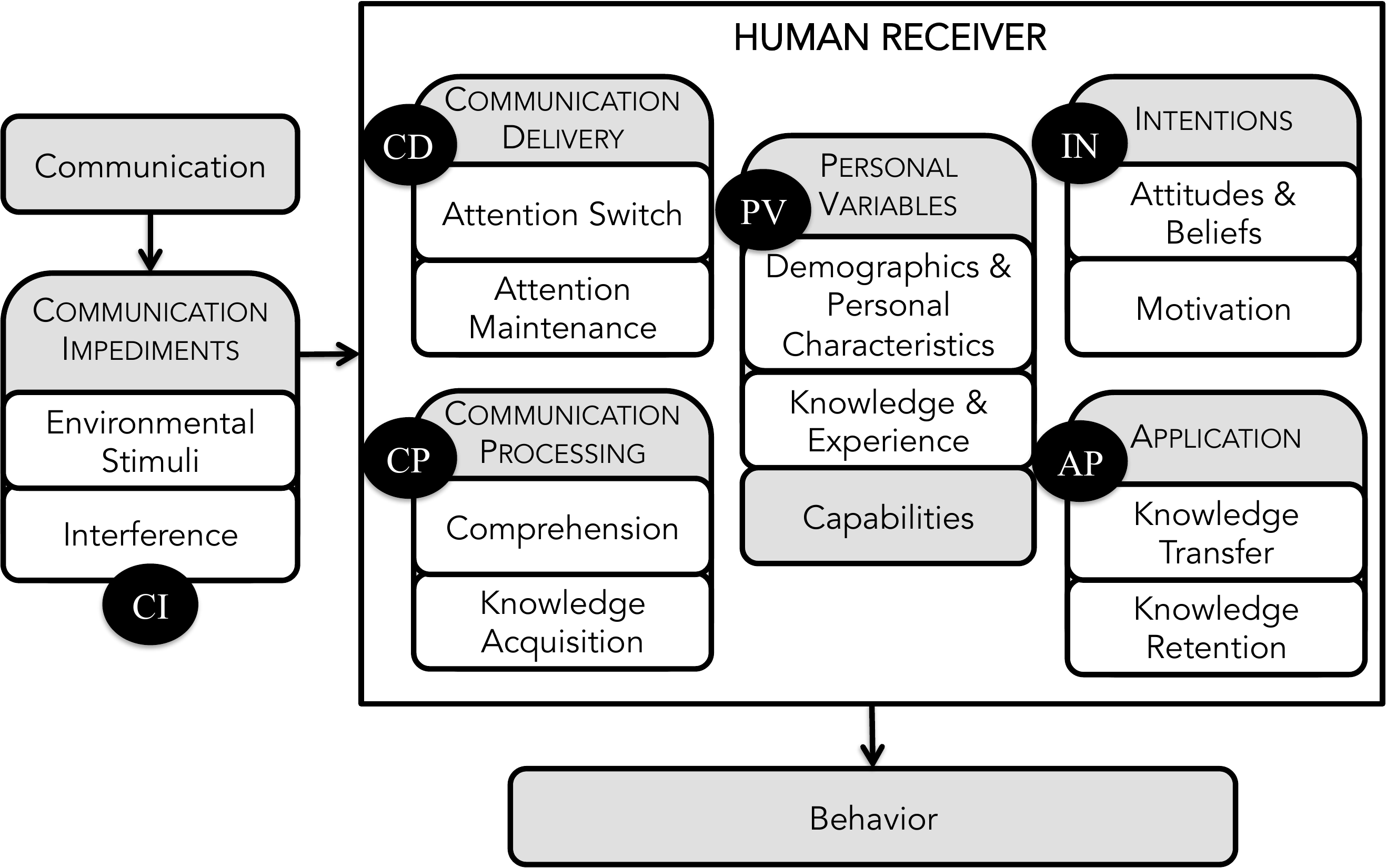}
	\caption{Cranor's Human in the Loop Security Framework \cite{cranor2008framework} ( Layout tweaked due to space constraints, and acronyms added for later reference).}
	\label{fig:chip}
\end{figure}

\subsection{Security vs. Privacy}
It is important to realise that security and privacy are fundamentally different concepts. 
Skinner \etal \cite{skinner2005framework} argue that a secure information system does not necessarily imply that privacy will be preserved in the system.
 Gritzalis and Lambrinoudakis \cite{gritzalis2008privacy} make the distinction between privacy and security as follows:  ``{\em a piece of information is secure when its content is protected, it is private when the identity of the owner is protected}'.
 As an example, they refer to a company that collects customer information and stores it encrypted. This ensures that the information is secured. Yet the same company may sell the information to another company, thereby violating the owners' privacy. 
 
 Bambauer \cite[p. 667]{bambauer2013privacy} explains: ``{\em Privacy discourse involves difficult normative decisions about competing claims to legitimate access to, use of and alteration of information}.''
 Security, on the other hand, is ``{\em the protection of information and information systems from unauthorized access, use, disclosure, disruption, modification, or destruction in order to provide confidentiality, integrity, and availability}.'' \cite{NIST}.
 
 Privacy and security are clearly distinct concepts, but their alerts still share some common characteristics in that they exist to tell the end user to something important. We therefore present three lists of guidelines: (1) generic, (2) privacy-specific, and (3) security-specific.

\section{Contemplating the Alert Literature}
\label{analysis}

We decided to focus on browser alerts firstly because of the popularity of web applications  \cite{mikowski2013single} such as email, claimed to be the most popular application in use \cite{alharbi2008graphical} and video streaming \cite{kellerman2010mobile}.
The second reason is that browsers run on all devices, ranging from Desktops to Smartphones. We felt that our guidelines could be maximally useful to developers if we focused on guidelines for browser alerts.

The literature search was carried out between November and December 2017 as follows: 

{\bf Databases:} ACM, Springer, Web of Science, Scopus, IEEE, and then Google Scholar to identify publications that did not appear in the databases.

{\bf Keywords:} `design guidelines' {\em and} `browser'{\em and} (`security' {\em or} `privacy) {\em and} (`feedback' {\em or} `warnings' {\em or} `alert' {\em or} `notification')

{\bf Time Range:} 2007---2017

 {\bf Exclusion Criteria:} Patents, citations, non-peer reviewed, not English or unobtainable.

\begin{table}[ht]
\begin{tabular}{l|>{\hfill}p{1.5cm}|>{\hfill}p{1.5cm}|>{\hfill}p{1.5cm}}
  Database   & Papers \newline Returned & Papers \newline Excluded & Papers \newline Analysed \\ \hline\hline
  Scopus   & 2 & 1 & 1\\
  \hline
  ACM & 12 & 9 & 3\\
  \hline
  Springer & 214 & 199 & 15\\
  \hline
  Web of Science & 0 & 0 & 0\\
  \hline
  Google Scholar & 181 &134&47\\
    \hline
  IEEE & 79 & 73 &  6\\
  \hline\hline
  Total & \multicolumn{3}{r}{72}\\
  \hline\hline
\end{tabular}
\end{table}

\subsection{Quantitative analysis}
\begin{figure}[ht]
	\centering
        \includegraphics[width=\columnwidth]{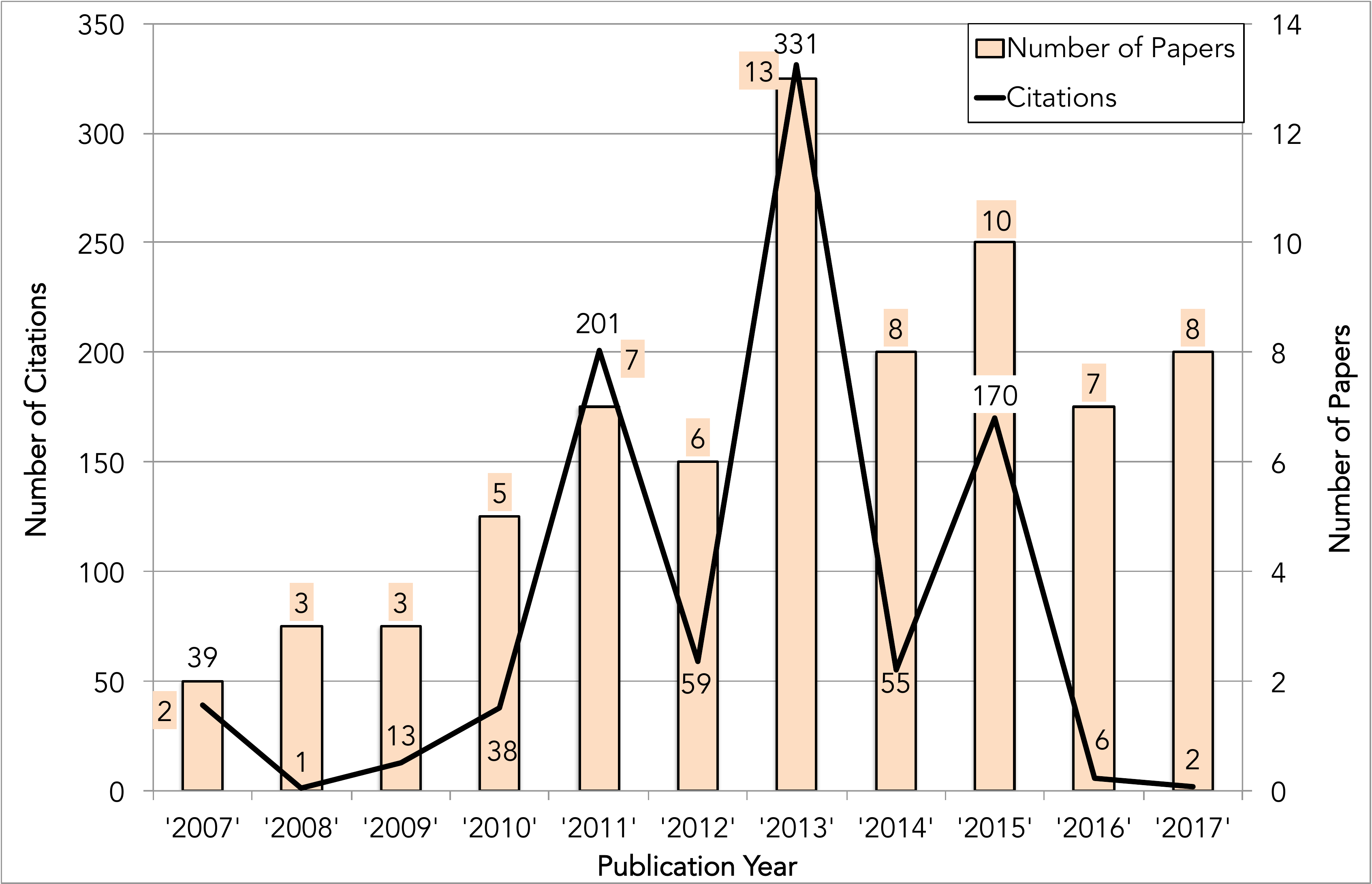}
	\caption{Number of Papers and Citation Numbers per Year}
	\label{fig:chart1}
\end{figure}

One particular measure of activity in a research field is the number of papers published over the decade in question. Figure \ref{fig:chart1} shows the number of papers, and also how many times the papers have been cited up to the date we carried our our literature review. 

It is interesting to note that 25 of the 72 papers had no citations at all. The average number of citations is 7.38, the mode is 0, and the median is 2. Only four of the papers had been cited by more than 50 other publications.
The top two most-cited publications appeared in conferences and the third most-cited publication appeared in a journal. 

\begin{figure}[ht]
	\centering
        \includegraphics[width=\columnwidth]{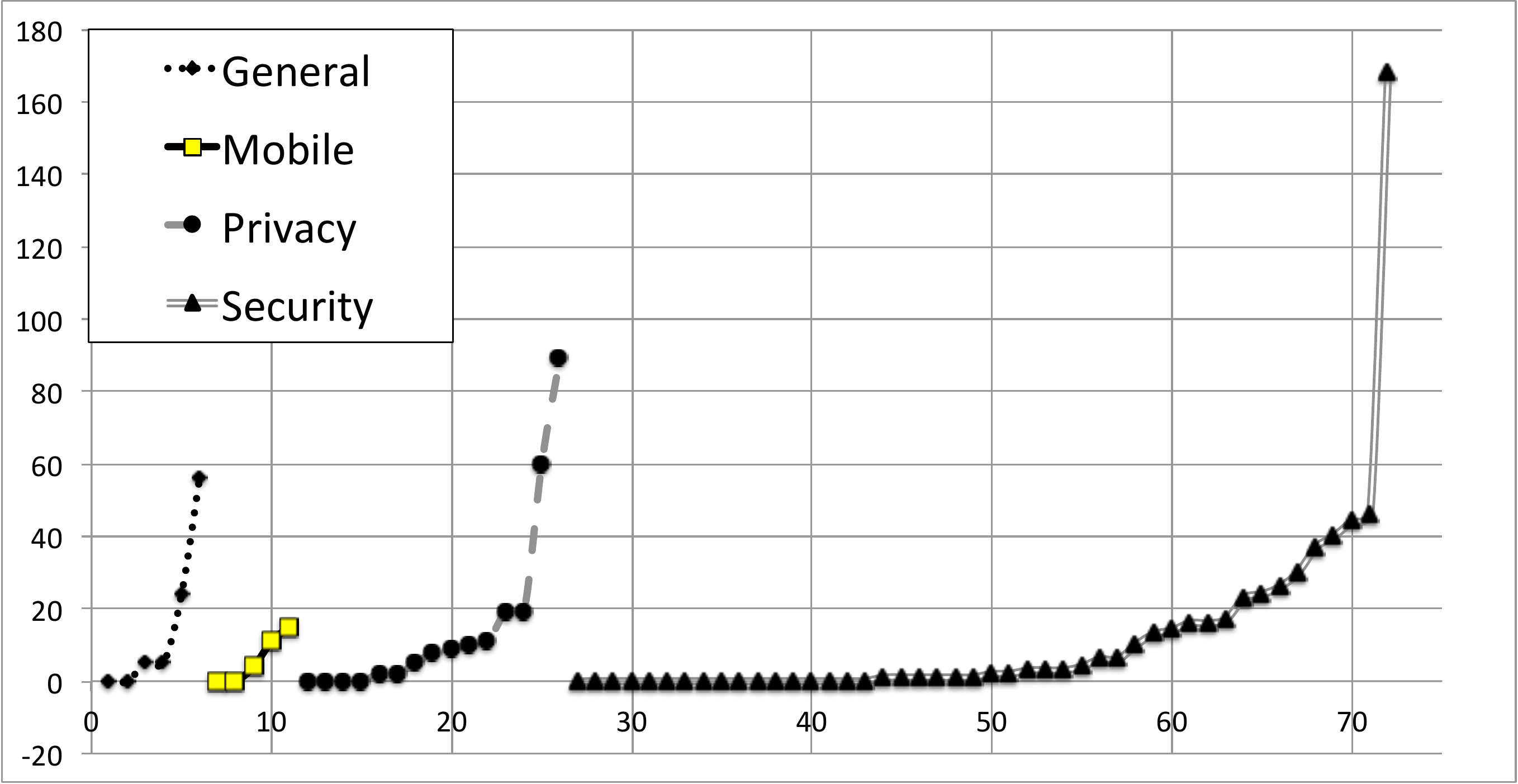}
	\caption{Number of Citations per Paper (by paper focus)}
	\label{fig:chart3}
\end{figure}

Figure \ref{fig:chart3} shows the citations for papers in each of the paper focus areas. The top cited paper is a security paper, with the next two most-cited papers being in the privacy area.

\begin{figure}[ht]
	\centering
        \includegraphics[width=0.8\columnwidth]{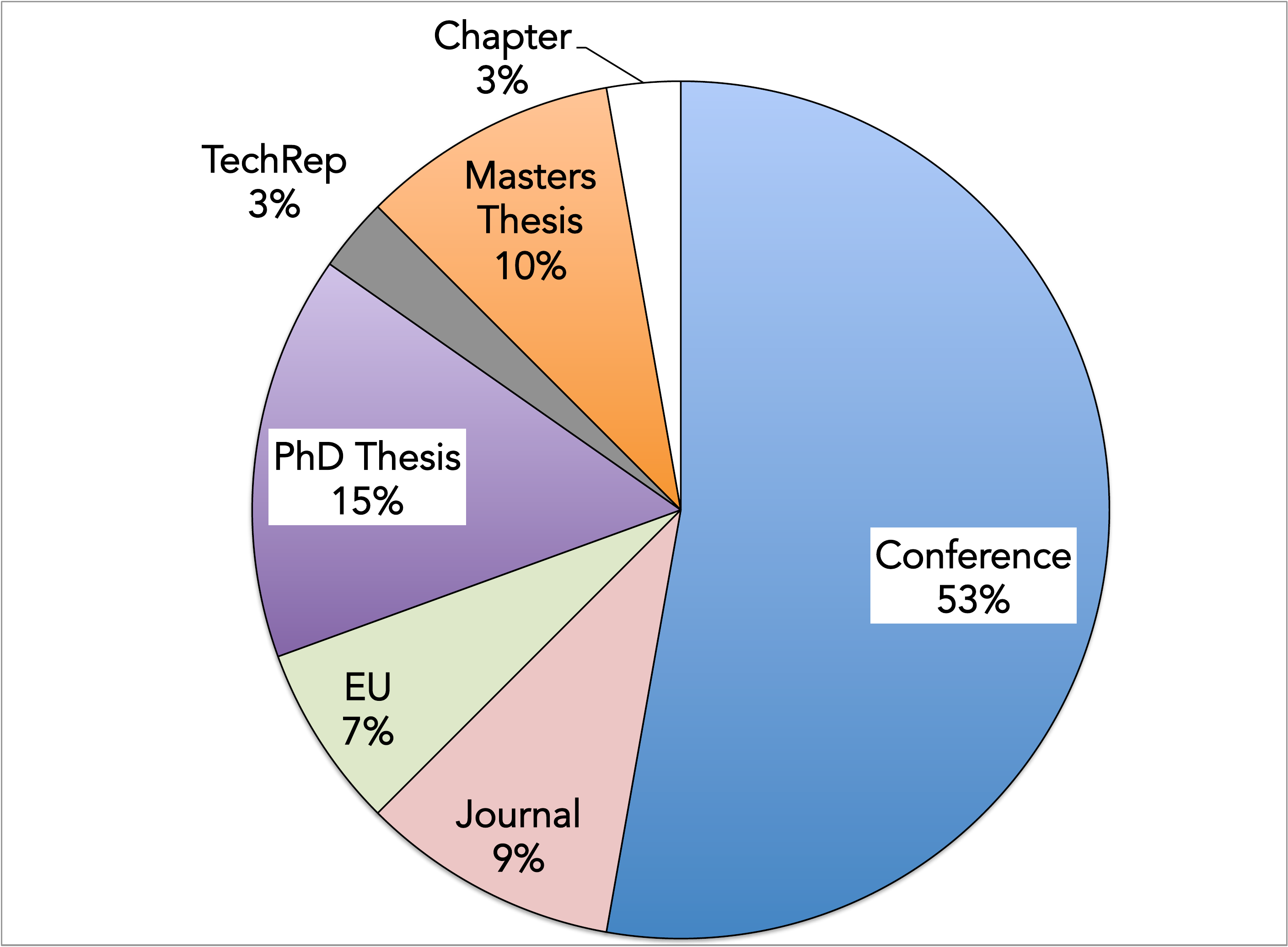}
	\caption{Publication Types}
	\label{fig:pie}
\end{figure}

Figure \ref{fig:pie} demonstrates the nature of the references we found. It is interesting that so many of the guidelines appear in Masters and PhD theses (18). Of these, 10 were never cited. The most-cited thesis, a PhD, was cited 13 times. Eight of the 10 PhDs had never been cited. The average number of citations across all theses was 2.47, but the mode and median are both 0. This suggests that  guidelines published in these formats have not made a significant impact on the field.   

\subsection{Qualitative analysis}
We analysed the guidelines using Thematic Analysis \cite{guest2012introduction}. This approach supports pinpointing, examining, and recording themes that emerge from the papers. We commenced by familiarising ourselves with the papers. We then generated initial codes and searched for themes as we collated these codes. We then reviewed the themes, defining and naming them.  
 
Some of the guidelines applied equally to privacy and security, but others were clearly specific to
either privacy or security. This is not unexpected because, as argued earlier,
 privacy and security are fundamentally different concepts. 
 
 We shall thus present generic guidelines first, then concept-specific guidelines separately, reflecting the fundamental differences between the two concepts.

\subsection{Generic Guidelines}\label{report}
We report first on the generic themes that coincide with Cranor's framework \cite{cranor2008framework}, depicted in Figure \ref{fig:chip}.

\subsubsection*{Communication Impediments (CI)}
    Here  mitigations to ameliorate the effects of impediments should be included.
   For example, provide users with the means to recover from hasty decisions \cite{PRIME}.
   
\subsubsection*{Personal Variables \& Capabilities (PV)}   
Some users may have low numeracy levels. Instead of providing them with figures regarding risk, perhaps utilise emotions or mood.  Similarly, users may have different understanding of visuals  \cite{nursespi2013}. Only by testing can the efficacy of these be confirmed. 

\subsubsection*{Communication Delivery (CD)}

   Human attention is a finite resource
\cite{colnago2016privacy}. Do not squander it, and do not expect the recipient to give you any as a matter of course.
   
    \textbf{\emph{CD1: Modality ---}} Murphy-Hill \& Murphy \cite{murphy2014recommendation,Westermann17} suggest that pictures be used to ease communication. When delivering warning alerts, users prefer text and graphical-based information, rather than aud\-it\-ory in\-for\-ma\-tion\cite{Chen:2011:IDE:2008579.2008601}.
    
    On the other hand, Goldberg \cite{goldberg2009state} suggests that text should be used exclusively. This might be to maximise accessibility, and the middle road would be to ensure that alt text is provided for all images. 
    
    Work conducted by Anderson \etal \cite{anderson2015polymorphic, anderson2016warning} notes the use of polymorphism in warning alerts to reduce habituation.

      \textbf{\emph{CD2: Timing ---}} If alerts appear too often the recipients may get annoyed  and start ignoring them \cite{murphy2014recommendation,akhawe2013alice}. Alerts should be issued only when necessary, to avoid irritating the user \cite{DBLP:conf/trust/VolkamerRCRB15}.

Westermann \cite{Westermann17} found that people were most annoyed by alerts if they are busy with a task, especially when the task is complex, but less annoyed in between tasks or when they were waiting for something.

It is important to prioritise the warnings so that only the most important ones merit immediate interruption \cite{Westermann17,akhawe2013alice}.

 \textbf{\emph{CD3: Location ---}} Westermann  \cite{Westermann17} considers alert location to be important. Many browsers, for example, display alerts either in the address bar (padlock and the word Secure in Chrome), or at the bottom of the screen. These are easily missed by users. If we want people to notice the alerts it ought to be displayed where they are currently focusing their attention.  In particular \cite{akhawe2013alice} point out that passive toolbar-located warnings are less effective than full page warnings that users are less likely to miss.
 Pala and Wang\cite{Pala:2009:UUI:1927830.1927852} also suggest alerts should be placed where the users are focusing their attention.
 In the study conducted by Chen, Zahedi, and Abbasi, users preferred alerts to be placed in the centre of the screen \cite{Chen:2011:IDE:2008579.2008601}.

 \textbf{\emph{CD4: Appearance ---}}  Kelley \cite{kelley2009designing} provides a number of recommendations: (1) the alert should be surrounded by a box to clearly demarcate it;
 (2)  provide a title to assist speedy recognition.
. 
     Be careful with colour use so as not to disadvantage those with colour deficiencies \cite{goldberg2009state}.
A neutral grey colour can be used for the background of alerts, as it is unlikely to annoy the user \cite{DBLP:conf/trust/VolkamerRCRB15}.

\subsubsection*{Communication Processing (CP)}
\ \\
 \indent \textbf{\emph{CP1: Make Essential Information Pertinent ---}} Lin \cite{lin2013understanding} suggests highlighting the most important information.  Keep initial details about the risk to a minimum \cite{nursespi2013, nurse2011past}.  Only the most important information should be displayed to the user immediately with links to more information should they want it \cite{DBLP:conf/trust/VolkamerRCRB15}. The granularity of information is important. Too-wordy information will not be read, and information that is too condensed can be obscure. In providing  alerts and alerts a balance must be found \cite{balebako2013little}.

 \textbf{\emph{CP2: Maximise Understandability ---} } \cite{kelley2009designing,murphy2014recommendation,Westermann17,shah2016evaluating} and consistency \cite{almeidamerging,murphy2014recommendation}.
Provide concrete explanations \cite{ozok2010design}. The importance of this aspect is confirmed by \cite{Ng2017}. Keep explanations simple \cite{lin2013understanding}.
Acronyms and jargon should be avoided and the use of meaningful terminology encouraged \cite{balebako2013little,kelley2009designing,shah2016evaluating,silic2015warning}. Separate semantically different kinds of information \cite{kelley2009designing,vasalou2015understanding}.

 Text presented should be easy for users to comprehend \cite{DBLP:conf/trust/VolkamerRCRB15}.  Short, simple sentences, devoid of complex grammatical structures should be used.  The use of technical words should be avoided (i.e. words listed in the indexes of IT security books)\cite{harbach2013sorry, Harbach:2012:TMW:2382196.2382301, Pala:2009:UUI:1927830.1927852, DBLP:journals/telsys/DongCJ10}.  Unclear alerts are more likely to be ignored, and consideration should be given to the exact meanings of words used \cite{nursespi2013}.

Work by Bravo-Lillo \etal.  investigated the use of redesigned warning alerts.  Longer warning alerts performed poorly in user testing, suggesting users may have become confused \cite{DBLP:conf/interact/Bravo-LilloCDKS11}.
Although existing work highlights shorter alerts are most effective at communicating security warnings to the user, the challenge of delivering such alerts whilst providing the user with an understanding has been acknowledged \cite{felt43265}.

\subsubsection*{Application (AP)}
\ \\
 \indent \textbf{\emph{AP1: Be Specific ---}}
Bravo-Lillo \etal \cite{DBLP:conf/interact/Bravo-LilloCDKS11} state that \textit{``to be successful, warnings should both motivate a user to respond, and help users understand the risk, in that order''}.

 Always tell the users what actions to take, if indeed they should take 
action. 
\cite{almeidamerging,PRIME}.

\textbf{\emph{AP2: End Goal ---}} Consider the way in which you want to communicate a  risk to the user e.g. is the alert to draw them away from a risky situation, or is the alert to help users understand the risk \cite{nursespi2013}?

\textbf{\emph{AP3: Effort does not Deter ---}}  Akhawe and Felt \cite{akhawe2013alice} explain that designers should not use the number of clicks required to bypass a warning to deter users. Their study showed that users were not sensitive to the number of clicks once they had made a decision.

\subsubsection*{Intentions (IN)}
\ \\
 It is important to note that delivering warnings is worthwhile. Silic \etal \cite{silic2015warning} found that people definitely took note of displayed warning  messages, suggesting that they thought about the information before making the decision to proceed. If people are reading and thinking about messages, these messages have a chance of changing attitudes and beliefs.

Vasalou \cite{vasalou2015understanding} says alerts should give recipients ``space for interpretation'', so that they can interpret the information as it applies to themselves personally. 

Phrasing of alerts could be personalised, depending on the skill level of the user, and their experience \cite{Chen:2011:IDE:2008579.2008601, nursespi2013, nurse2011past}.  Personalised alerts were said to be successful when utilised to inform users about two-factor authentication, and bullet-points can be used to aid clarity of information presented \cite{redmiles2017}.

 It is important for the user to retain a level of control \cite{PRIME, vasalou2015understanding,xu2012value}.
 Schaub \etal \cite{schaub2015design} distinguish between three levels  of user control: (1) blocking, non-blocking and decoupled. 
 A designer has to
 decide whether the user has to acknowledge the message (blocking) or not (non-blocking), whether they can defer it (decoupled), or whether it will expire after displaying for a certain period of time \cite{murphy2014recommendation}.

 Users should be provided with the option to respond to a  risk they have been alerted to, and helped to visualise potential consequences \cite{nursespi2013}.  Work by Volkamer \etal \cite{DBLP:conf/trust/VolkamerRCRB15} concurs that the potential consequences of a  risk should be conveyed to the user, along with potential recommendations.

Make sure the user can easily get in touch with someone to ask about warnings \cite{goldberg2009state}. Information should be conspicuously placed so worried users will be able to get help  \cite{PRIME}.

\subsection{Privacy-Specific Guidelines}

Allow users to make privacy choices that are  (1) meaningful, (2) informed, (3) timely  \cite{colnago2016privacy}.

\textbf{\emph{P-CI: Inspire Trust ---}} Trust should be deliberately built and maintained \cite{murphy2014recommendation} by framing the privacy alert very carefully \cite{adjerid2013sleights}.
Rather counter-intuitively, privacy alerts should not provide justifications for information requests. Researchers report that justifications potentially reduce the end-user's trust in the system \cite{aagaard2013privacy,adjerid2013sleights,Knijnenburg15,pollach2007s}.

\textbf{\emph{P-PV: Privacy Expectations ---}} Lin \cite{lin2013understanding} points out that users have different privacy expectations, and that an alert interface should reflect this reality.

\textbf{\emph{P-CD: Specificity ---}} Ensure that the sensitivity of the data is communicated go to the user \cite{nafra14}.

\textbf{\emph{P-IN: Enhance Control ---}} Ensure that control resides with the user \cite{nafra14}. Do not merely report that some privacy invasion has occurred: allow the user to control disclosure.
 It is necessary to balance interruptions and ensuring that the user retains a sense of control \cite{colnago2016privacy}.
 
People have different levels of privacy concerns, and the alerts should afford users the level of control matching their personal privacy concern.
 
\subsection{Security-Specific Guidelines}

Herzog and Shahmehri highlight the importance of security features in applications, stressing that \textit{``security is rarely the primary user task''} \cite{Herzog:2007:UHT:1234772.1234787}.

\textbf{\emph{S-CI: Context-Sensitive Help ---}} Constantly visible context sensitive help may prove useful in helping the user understand security. Help may be provided via the use of an agent \cite{Herzog:2007:UHT:1234772.1234787}.  The user should be provided with the option to find  further information in a contextually-aware setting \cite{Pala:2009:UUI:1927830.1927852, nurse2011past}.

\textbf{\emph{S-CD1: Provide Justification ---}} The user needs to know why the alert is being provided \cite{murphy2014recommendation}. 

Provide information as to whether a component is secure or insecure.  By displaying this information in either case, this provides a consistent interface for the user \cite{Pala:2009:UUI:1927830.1927852}.
Ensure the current state of the system is displayed to the user \cite{nurse2011past}.

\textbf{\emph{S-CD2: Colour ---}}
 
Research regarding two-factor authentication suggests the use of blue as a peaceful colour.  Red might indicate an incident has occurred \cite{redmiles2017, Chen:2011:IDE:2008579.2008601}.
Felt \etal \cite{felt43265} suggest utilising \textit{``opinionated design''}.  For example,  make the ``correct'' response the more visually appealing option e.g. the button should have a high contrast level against the background.
Others have utilised green as colour, noting that it is seen as safe.  Whilst users should be given options regarding how to proceed with their tasks, it has been suggested that placing the ``correct'' option in green serves to guide users towards the safe choice \cite{DBLP:conf/trust/VolkamerRCRB15}.

Where colourblind users may have issues with warnings, the use of secondary information (icons) aims to convey the same message \cite{DBLP:conf/trust/VolkamerRCRB15}.

\textbf{\emph{S-CD3: Graphics ---}} 
In one study, participants felt the inclusion of graphics in an alert about two-factor authentication conveyed a tone which was less serious, and suspicious \cite{redmiles2017}.
Conversely, other studies conclude graphics are required in alerts, to convey reassurance, draw attention, and to reduce cognitive effort \cite{DBLP:conf/trust/VolkamerRCRB15}.
This is a prime example of a conflicting set of guidelines. 

Eargle \cite{eargle31614} suggests that facial expressions could be used to convey  threat levels in security alerts but this has not been confirmed by any other studies in our studied group. 

\textbf{\emph{S-IN: Control Level ---}} If a security issue is detected on a page, users would prefer the security alert to block them from visiting a malicious website \cite{Chen:2011:IDE:2008579.2008601}.
Other research stated the final security decisions should be left to the user, though users should be provided with alternative options on how to proceed with their task \cite{DBLP:conf/trust/VolkamerRCRB15}.

\section{Informing Designers}
\label{examples}
\begin{figure}[ht]
	\centering
        \includegraphics[width=\columnwidth]{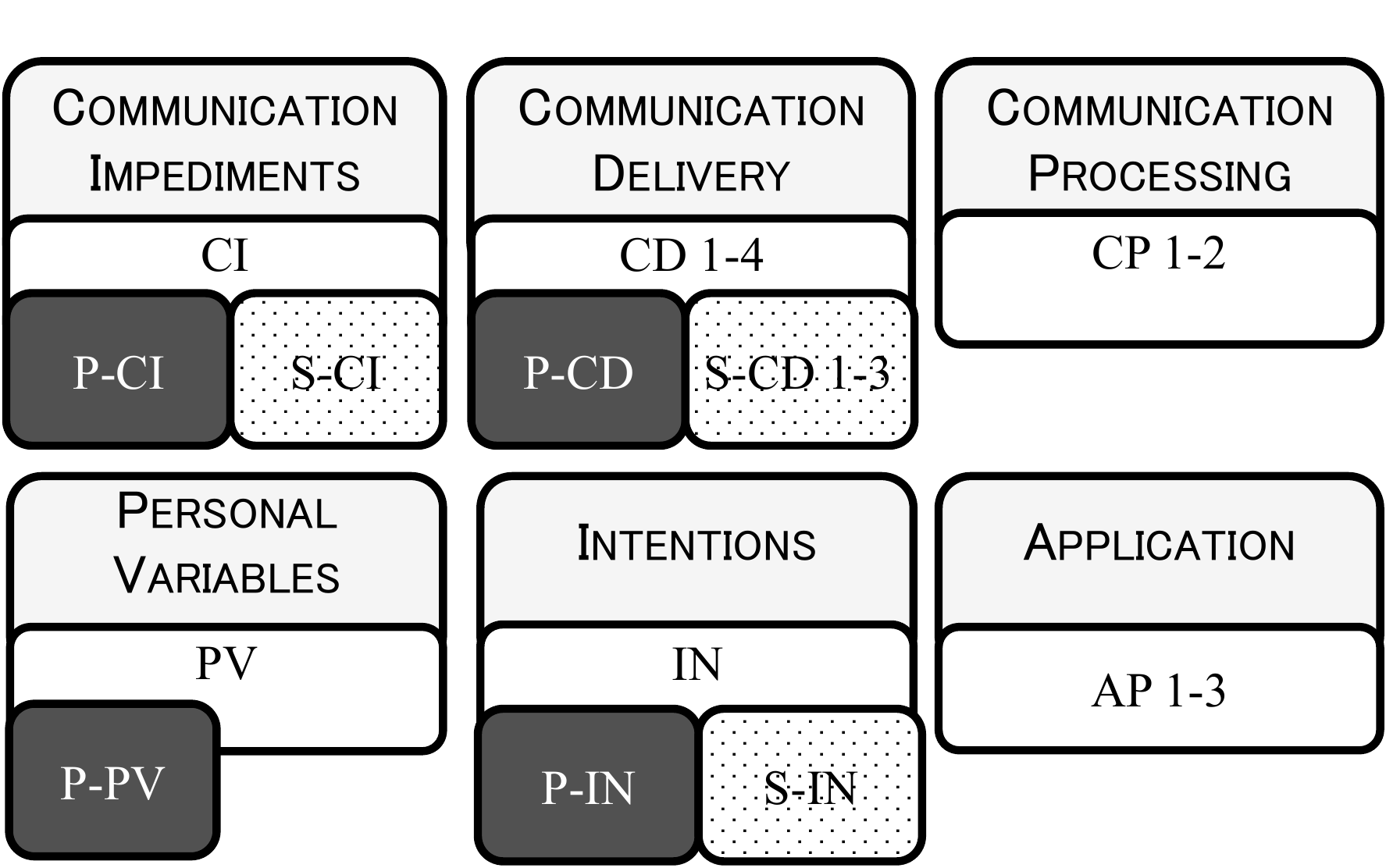}
	\caption{Consolidated Guidelines}
	\label{fig:list}
\end{figure}

The previous section provided a list of recommendations for designing alerts (Figure \ref{fig:list}).
However, as pointed out by \cite{Renaud17}, and confirmed by \cite{luger2014value},
designers have difficulty benefiting from these kinds of flat lists of guidelines.

\begin{figure}[ht]
	\centering
        \includegraphics[width=\columnwidth]{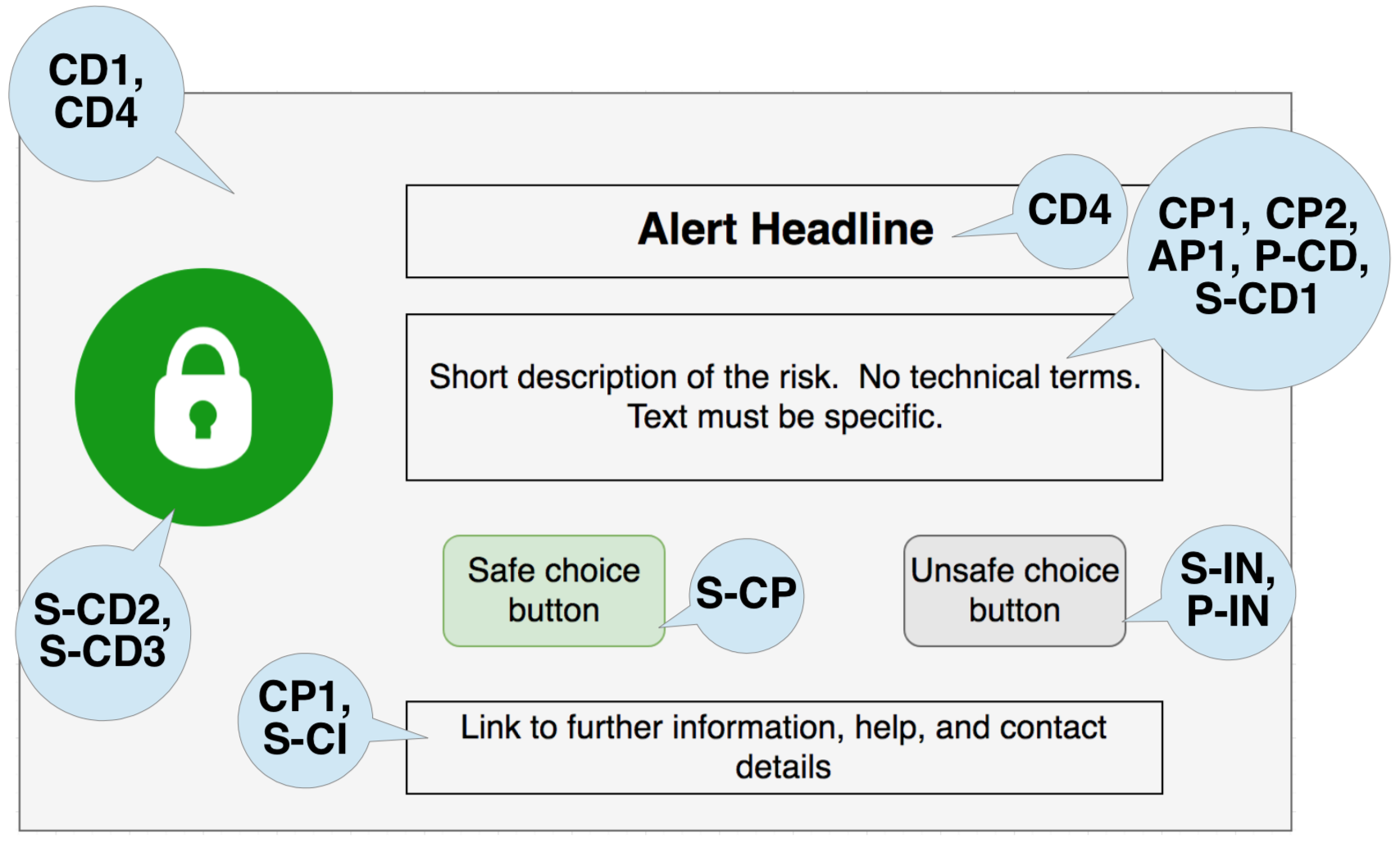}
	\caption{Example alert template}
	\label{fig:templatealert}
\end{figure}

Luger and Rodden's \cite{luger2014value}
designers spoke about the value of examples in encapsulating the lessons of design guidelines in a more palatable format. We will thus present a template example  of an alert for the benefit of designers, extending the initial template produced by \cite{Bauer13}. We thus present an alert template in Figure \ref{fig:templatealert} and explain how it satisfies the guidelines.

\textbf{Generic Guidelines: }
The template contains both an icon, and text to communicate the contents of the alert (CD1).  The alert has a border and a headline title, along with the use of colours and icons.  The background colour of the template is neutral (CD4).  The text explaining the alert should be clear, specific, and easy to understand, requiring minimal cognitive effort (CP1, CP2, AP1).  If a user would like to find out more information, they should be presented with the opportunity to access this, along with relevant contact details.%

\textbf{Privacy Guidelines: }%
If the alert is being used to notify the user about privacy, the sensitivity of the information being dealt with should be clearly communicated to the user (P-CD).  Users are provided with a choice in the alert, ensuring they remain in control (P-IN).%

\textbf{Security Guidelines: }%
The text explaining the alert should justify why it is being displayed (S-CD1).  The safe choice button on the template alert is more visually appealing than the unsafe choice, and it clearly contrasts with the background of the alert (S-CP).  Users are provided with a choice in the alert, ensuring they remain in control (S-IN).  Users should be presented with the option to access context-sensitive help.  Colour and graphics should be used to aid in communicating the role of the alert, ensuring colourblind users are not placed at a disadvantage (S-CD2, S-CD3).%

\textbf{Template Summary: }
Graphics and text are used to communicate the nature of the alert (CD1, CD4, S-CD2, S-CD3).  A headline title and a neutral background are used (CD4). The text explaining the alert should be clear, specific, and easy to understand, requiring minimal cognitive effort (CP1, CP2, AP1, PC-D, S-CD1).  Users should have the opportunity to access further information, and relevant contact details (CP1), along with context-sensitive help (S-CI).  Users are provided with a choice in the alert, ensuring they remain in control (P-IN, S-IN).  The safe choice button is more visually appealing than the unsafe choice, clearly contrasting with the background of the alert (S-CP).

\subsection*{Development Good Practice}

Creating a well-designed environment can aid in establishing trust  \cite{mendoza2015}.
Moreover, it is important to ensure that people are receptive to alerts \cite{Westermann17}. The best way to confirm both trustworthiness and alert receptiveness is by means of thorough testing \cite{PRIME}. Options are A/B testing in the wild, controlled experiments, field studies \cite{akhawe2013alice}, or case studies post-deployment \cite{murphy2014recommendation}.

\section{Future Work}
\label{futurework}
The systematic literature review identified a lack of research surrounding the optimal placement of security and privacy alerts within a the web browser.  Whilst work carried out by Chen, Zahedi, and Abbasi \cite{Chen:2011:IDE:2008579.2008601} showed users preferred alerts in the centre of the screen, usability studies have shown there are a variety of patterns users exhibit when browsing web content \cite{pernice_2017}.  This suggests that further research is required into the optimal placement of security and privacy alerts.

It is also interesting to note from Figure \ref{fig:list} that there are no security or privacy-specific guidelines in terms of Communication Processing or Application. These are certainly areas for further investigation.

Several guidelines gathered from literature conflict, and this  issue has been highlighted by other guideline papers \cite{Renaud17}.  Previous research has acknowledged that \textit{``Not all best practices can be simultaneously satisfied''}; therefore, trade-offs must occur \cite{felt43265}.  Masip \etal \cite{DBLP:conf/hcse/MasipMWPGO12} have investigated the development of a design process to assist with design choices when there are potentially conflicting user interface guidelines.  In the future, we plan to develop a methodology for prioritising the guidelines to support security and privacy alert design.

\section{Conclusion}
\label{conc}
The systematic review process highlighted a large proportion of the work found online relating to alerts were sourced from student theses (both at Masters and PhD level).  Whilst conducting the analysis process, it became clear that some alert guidelines were developed for security, and others were developed for privacy. These seemed, in many cases, to be fundamentally different, suggesting that different guidelines are required for these two distinct areas.  

We publish this work as a first attempt to provide guidance to designers and developers who need to incorporate alerts into their systems.  In the future, we seek to prioritise the guidelines, addressing the issue of potential conflicts, and with feedback from practitioners, iteratively refine the guideline list.

\balance
\bibliographystyle{ACM-Reference-Format}
\bibliography{refs}

\end{document}